\documentclass[11pt]{article}
\usepackage{moriond,epsfig}

\def\Journal#1#2#3#4{{#1} {\bf #2}, #3 (#4)}

\def\NPB{{\em Nucl. Phys.} B}

\def\PRL{\em Phys. Rev. Lett.}

\def\be{\begin{equation}}
\def\ee{\end{equation}}
\def\bea{\begin{eqnarray}}
\def\eea{\end{eqnarray}}

\def\NP{\em Nucl. Phys.}

\begin{document}
\vspace*{4cm}
\title{SEARCHES FOR CHARGED HIGGS BOSONS AT LEP}

\author{ Andr\'e Holzner }

\address{Institut f\"ur Teilchenphysik, ETH H\"onggerberg,
  CH-8093 Z\"urich, Switzerland}

\maketitle\abstracts{
The four LEP experiments Aleph, Delphi, L3 and Opal updated their searches for pair production of  
charged Higgs bosons using more than $210\ \mathrm{pb}^{-1}$ luminosity
collected per experiment in the year 2000. Combining it with previously collected
data, a significant deviation from background (equivalent to 4.4 $\sigma$) 
is found by the L3 collaboration for low values of the branching ratio
$\mathrm{H}^\pm \to \tau\nu$ around masses of about 68\ GeV. This excess
is however not seen by the other LEP collaborations and thus a lower limit
on the charged Higgs mass is set at 78.5 GeV at 95\% confidence level.
All results reported here are still preliminary.
}

\section{Theoretical Framework}

  In the Standard Model~\cite{sm}, the particle masses are generated
  through the coupling to the Higgs field. The Standard Model is
  minimal in the sense that it contains only one Higgs doublet
  which corresponds to four degrees of freedom. Three out of these
  area `eaten up' by the longitudinal polarization states of the W and
  Z bosons, the remaining one is the physical Higgs field. However, 
  there is no a priori reason why there should only be one such
  doublet. Two Higgs Doublet Models (2HDMs)~\cite{higgs-hunter} have eight degrees of freedom
  coming from the Higgs sector. Five of these correspond to the Higgs particles out of which two are 
  the charged Higgs bosons $\mathrm{H^+}$ and $\mathrm{H^-}$. Their masses are predicted to be equal. 

  The charged Higgs sector of 2HDMs can be parameterized by the mass of
  the charged Higgs boson $m(\mathrm{H}^\pm)$ and $\tan\beta$ which
  is the ratio of the vacuum expectation values of the two doublets.

  Two types of 2HDMs are distinguished: The absence of flavor changing neutral
  currents puts a lower limit on the charged Higgs mass in e.g. type
  II models without further particle content~\cite{ciuchini}, while type I models are 
  much less restricted by experimental data. For low values of $\tan\beta$, the
  dominant decay modes are $\mathrm{H^+} \to \mathrm{c\bar{s}}$ and $\tau^+\nu_\tau$
  (and charge conjugates) while for high values values of $\tan \beta$ the three body decay
  $\mathrm{H}^\pm \to \mathrm{W^{\pm*} A}$ dominates. Of the LEP experiments, only Opal has performed a search for three
  body decays of charged Higgs bosons~\cite{opal-moriond}.

  This document reports on the searches for pair production of 
  $\mathrm{H^+}$ and $\mathrm{H^-}$ in $\mathrm{e^+e^-}$ collisions at LEP. 
The dominant process is $\mathrm{e^+e^-} \to \mathrm{Z}^* / \gamma^* \to \mathrm{H}^+\mathrm{H}^-$.
  
\section{Experimental signatures and event selection}
The branching ratio $\mathrm{BR(H^\pm \to \tau\nu)}$ and the charged
Higgs mass are not predicted by theory and therefore left as parameters of the
search. It is assumed that $\mathrm{c}\mathrm{\bar{s}}$ and
$\tau\nu_\tau$ are only decay modes of the charged Higgs boson,
leading to three different final states: $\tau^+\nu_\tau\tau^-\bar{\nu}_\tau$
(leptonic channel),
$\mathrm{c}\mathrm{\bar{s}}\tau^-\bar{\nu}_\tau$~\footnote{The charge
  conjugate reaction is implied throughout this letter}
(semileptonic channel)
 and $\mathrm{c}\mathrm{\bar{s}}\mathrm{\bar{c}}\mathrm{s}$
(hadronic channel).

{\bf Leptonic Channel}: The leptonic channel is characterized by two acoplanar taus
(electrons, muons or low multiplicity jets) and missing energy.
No mass reconstruction is attempted since at least four neutrinos
are present in the final state. The most important backgrounds are $\mathrm{e^+e^-} \to \mathrm{W}^+\mathrm{W}^- \to
\ell^+\nu\ell^-\bar{\nu}$, $\gamma\gamma$ collisions and 
$\mathrm{e^+e^-}\to \tau\tau(\gamma)$ where the $\gamma$ escapes detection.

\begin{figure}
\begin{center}
  \psfig{figure=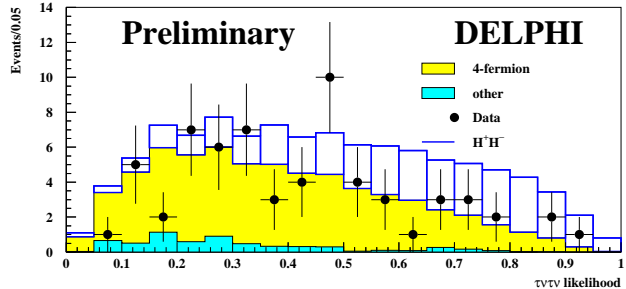,height=1.5in}
\end{center}
\caption{
  Likelihood variable of the DELPHI $\tau^+\nu_\tau\tau^-\bar{\nu}_\tau$ analysis.
  The shaded histograms represent the expected background, the open
  area is the signal expectation for a Higgs mass of 75 GeV (on top of the background). There is
  a good agreement between the observed data and the expected background.
   \label{fig:delphi-tntn-likelihood}
   }
\end{figure}

For example, Delphi uses a likelihood method to discriminate between
signal and background based on $\tau$ polarization variables
and the polar angle of the $\tau$ decay products~\cite{delphi-moriond}.
The distribution of the likelihood variable is shown in 
Figure~\ref{fig:delphi-tntn-likelihood}. No 
significant excess is observed.


\begin{figure}
\begin{center}
  \psfig{figure=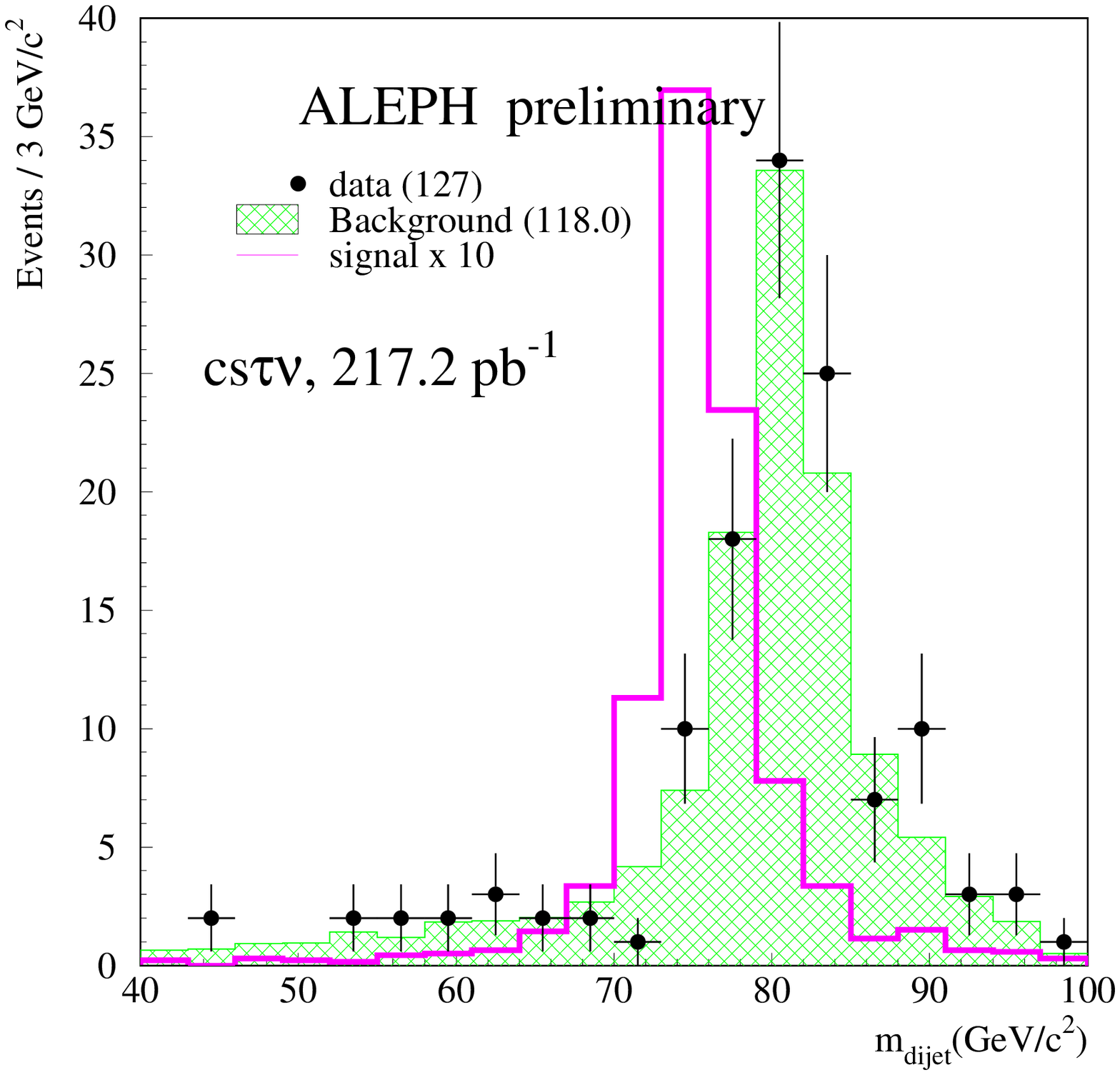,height=2.5in}\hspace{1cm}
  \psfig{figure=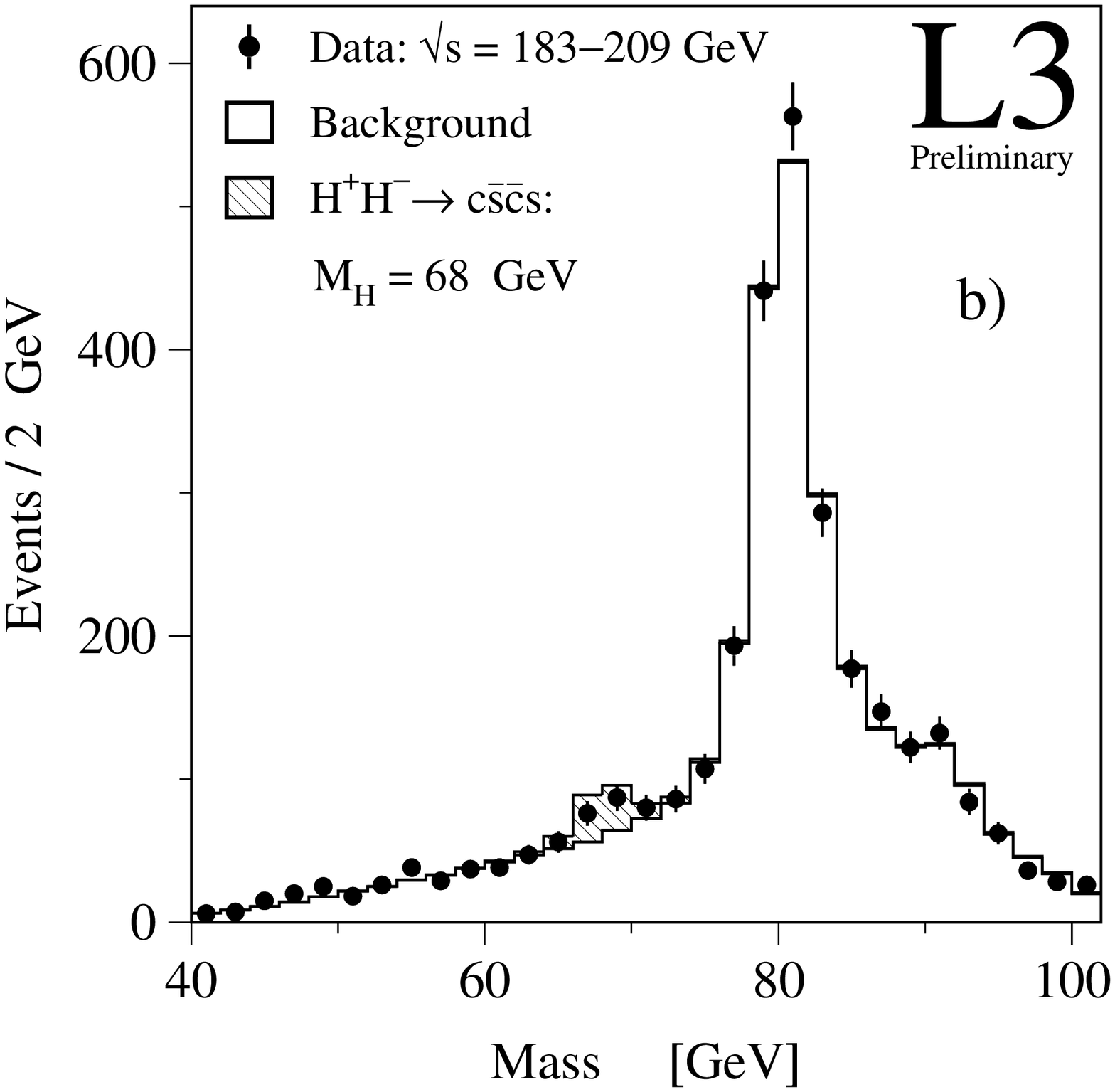,height=2.5in}
\end{center}
\caption{
  Left: Reconstructed Higgs candidate mass in the ALEPH $\mathrm{c}\mathrm{\bar{s}}\tau^-\bar{\nu}_\tau$
analysis.
  The hatched histogram is the expected background, the open
  histogram is  the expectation from a charged Higgs signal enlarged
  by a factor of ten for a Higgs mass of 75 GeV. Data is in good agreement with the
  background expectation. 
  Right: Reconstructed Higgs candidate mass in the L3 four jet analysis.
  The open histogram is the expected background, the hatched
  area is the expectation from a charged Higgs signal with
  a mass of 68\ GeV (on top of the background). 
  \label{fig:aleph-cstn-mass}
   \label{fig:l3-cscs-mass}
  }
\end{figure}

{\bf Semileptonic Channel:}
In this channel, high multiplicity events consistent with two jets plus an additional
electron, muon or low multiplicity jet are selected. The events
also must have a high amount of missing energy. 
The background is dominated by the process $\mathrm{e^+e^-}\to
\mathrm{W^+W^-} \to \mathrm{q\bar{q}}\tau\nu$.
As an example, Aleph uses linear discriminants in the analysis based on variables
like the $\tau$ decay angle in the Higgs rest frame, $\tau$ polarization probability,
Higgs production angle etc~\cite{aleph-moriond}. An example of the mass distribution after
a cut on the discriminant is shown in
Figure~\ref{fig:aleph-cstn-mass}a. Data favor the background-only hypothesis.

{\bf Hadronic Channel:}
The signature of this channel is a four jet structure and no missing
energy. The most important backgrounds are W pair and $\mathrm{q}\mathrm{\bar{q}}$
production. For example, L3 uses in their four jet analysis (among other cuts)
a neural network to discriminate $\mathrm{q}\mathrm{\bar{q}}$
events~\cite{l3-moriond}. The jets are grouped into pairs such that the probability
of the pairs having equal mass is maximized. A cut on the Higgs
production polar angle is applied to reduce the $\mathrm{W^+W^-}$ background.
The Higgs candidate mass distribution is shown in
Figure~\ref{fig:l3-cscs-mass}b. A significant excess is observed at
$m(\mathrm{H^\pm}) \approx 68$\ GeV.

\section{Combination of the Channels and LEP combination}

\begin{figure}
\begin{center}
  \psfig{figure=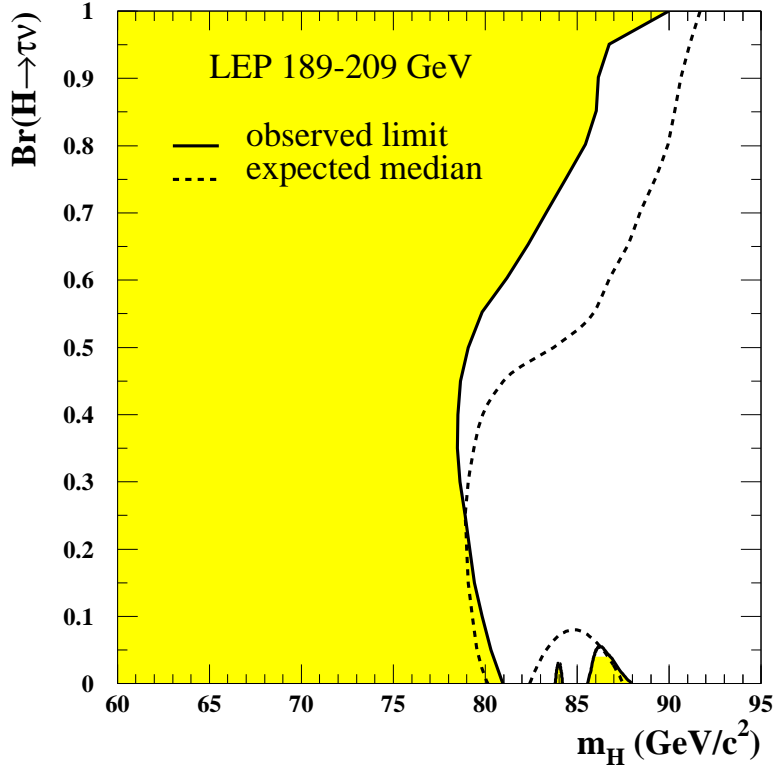,height=4.0in}
\end{center}
\caption{Excluded regions in the $m(\mathrm{H^\pm})$ vs. $\mathrm{BR(H^\pm \to
  \tau\nu)}$ plane for the combination of the four LEP experiments (shaded region).
  The dashed line shows the median expected limit (from background processes).
}
\label{fig:adlo-lim}
\end{figure}

Table~\ref{tab:limits} summarizes the observed and expected limits for fully hadronic and
fully leptonic branching ratio as well as the observed limits
independent of the branching ratio for the individual experiments
and their combination.
Figure~\ref{fig:adlo-lim} shows the excluded region
in the $m(\mathrm{H^\pm})$ vs. $\mathrm{BR(H^\pm \to
\tau\nu)}$ plane for the combination of the four LEP experiments~\cite{lephwg-charged-higgs}.


\begin{table}[t]
\caption{Observed (expected) lower limits on the charged Higgs mass in
  GeV at 95\% confidence level. The limits are shown for the four LEP
  experiments and their combination.
  \label{tab:limits}}
\vspace{0.4cm}
\begin{center}
\begin{tabular}{|c|c|c|c|c|c|}
\hline
                                    & Aleph       & Delphi      & L3          & Opal        & LEP\\ \hline
$\mathrm{BR(H^\pm \to\tau\nu)}$ = 0 & 80.7 (78.1) & 77.4 (77.0) & 77.2 (77.1) & 76.2 (77.1) & 81.0 (80.1) \\
$\mathrm{BR(H^\pm \to\tau\nu)}$ = 1 & 83.4 (86.9) & 85.4 (89.3) & 84.9 (83.0) & 84.5 (86.5) & 90.0 (91.7) \\
lowest observed                     & 78.0        & 73.8        & 67.1        & 72.2        & 78.5        \\ 
\hline
\end{tabular}
\end{center}
\end{table}


\section{Summary}

Searches for pair production of charged Higgs bosons in $\mathrm{e^+e^-}$
collisions were performed by the four LEP experiments.
A significant (4.4 $\sigma$ deviation from the expected background) 
excess is observed around $m_{\mathrm{H^\pm}} \approx 68\
\mathrm{GeV}$ in the L3 data over the last years, which is however not
seen by the other LEP experiments. Work is continuing to better 
understand the nature of this excess and the difference between the
observations of the four LEP experiments. Combining the L3 data 
with the other three LEP experiments, the deviation from background
is significantly smaller than what one expects from a charged Higgs 
boson of mass 68\ GeV, thus a lower limit of 78.5 GeV
is set on $m(\mathrm{H}^\pm)$ at 95\% confidence level. All results are
still preliminary.


\end{document}